\documentclass[aps,prl,twocolumn,superscriptaddress]{revtex4}
\usepackage{times,xspace}
\usepackage{amsbsy,amssymb,amsmath,bm}
\usepackage{graphicx,color,epsfig,rotate}
\begin{document}
%
\title{Spectroscopy of Magnetic Excitations in Magnetic
Superconductors Using Vortex Motion}
\author{L.N. Bulaevskii, M. Hru\v{s}ka, and M.P. Maley}
\affiliation{Los Alamos National Laboratory,
Los Alamos, New Mexico 87545}
\date{\today}

\begin{abstract}
In magnetic superconductors a moving vortex lattice is accompanied by
an ac magnetic field which leads to the generation of spin waves. 
At resonance conditions the dynamics of vortices in
magnetic superconductors changes drastically, resulting in strong
peaks in the dc I-V characteristics at voltages at which the washboard
frequency of vortex lattice matches the spin wave frequency
$\omega_s({\bf g})$, where ${\bf g}$ are the reciprocal vortex lattice
vectors. We show that if washboard frequency lies above the magnetic
gap, peaks in the
I-V characteristics in borocarbides and cuprate layered magnetic
superconductors are strong enough to be observed over the
background determined by the quasiparticles. 
\end{abstract}
%
\maketitle

The coexistence of magnetism and superconductivity was observed in
many crystals, such as RMo$_6$S$_8$, RRh$_4$B$_4$,
RBa$_2$Cu$_3$O$_{7-\delta}$ and (R,A)CuO$_{4-\delta}$ (A=Sr, Ce) with
the temperatures of magnetic ordering $T_M$ much smaller than the
superconducting critical temperature $T_c$, and also in borocarbides 
RT$_2$B$_2$C and ruthenocuprate RuSr$_2$GdCu$_2$O$_8$ with $T_M$ of the
same order as $T_c$. Here R is the rare
earth element, while T=Ni, Ru, Pd, Pt. In such crystals $f$-electrons
of ions R give rise to localized magnetic moments, while conducting
electrons exhibit the Cooper pairing. In all these crystals, except
HoMo$_6$S$_8$ and ErRh$_4$B$_4$, magnetic moments order
antiferromagnetically below $T_M$. Such magnetic ordering coexists
with superconductivity without strong interference because spin
density varies on the scale much smaller than the superconducting
correlation length and net magnetic moment vanishes, for review see 
Refs.~\onlinecite{rev1,bul,bul1}. 

In this Letter we consider interplay between magnetic and
superconducting excitations in magnetic superconductors, particularly
interaction between a moving vortex lattice and spin waves via the
ac magnetic field induced by moving vortices. 
The energy transfer from
vortices to the magnetic system leads to dissipation which is
additional to that caused by quasiparticles. This results in strong
current peaks in the dc I-V characteristics at voltages at which the
washboard frequency of vortex lattice \cite{Fiory} matches the spin
wave frequency $\omega_s({\bf k})$ and ${\bf k}$ matches a reciprocal
vortex lattice vector ${\bf g}$. 

First we consider slightly anisotropic
superconductors, i.e. all systems mentioned above except 
SmLa$_{1-x}$Sr$_x$CuO$_{4-\delta}$ and RuSr$_2$GdCu$_2$O$_8$ crystals,
and probably also 
Sm$_{2-x}$Ce$_x$CuO$_{4-\delta}$. The latter are layered
superconductors with intrinsic Josephson junctions 
\cite{Shibata,Ronning,Nachtrab,Cho}. 
 
We assume, for simplicity, a uniaxial crystal structure with the
principal axis along $z$. The dc magnetic field is applied along the
$z$-axis and we assume that the magnetic induction ${\bf B}({\bf r})$,
${\bf r}=x,y$, inside the superconductor corresponds to the ideal
Abrikosov square vortex lattice (such a lattice is realized in clean
borocarbide crystals in field ${\bf B}\parallel c$ in some field
intervals \cite{rev1}). The sublattice magnetization in the case of
antiferromagnetic ordering is assumed to be oriented in the $(x,y)$
plane. The dc transport current with the density ${\bf j}$ is along
the $y$-axis which, due to the Lorenz force, causes motion of the vortex
lattice with the velocity ${\bf v}$ along the $x$-axis.

\begin{figure} 
 \begin{center}  
 \epsfig{file=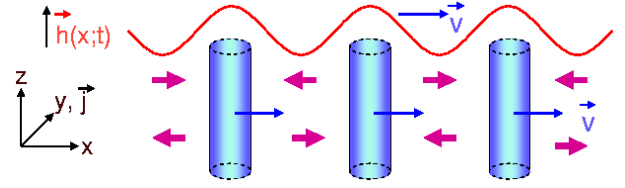,width=0.15\textwidth,clip=,angle=-90} 
 \end{center} 
 \caption{Vortex lattice moving with the velocity ${\bf v}$ induces
a spatially periodic ac magnetic field $h(x,t)$ which excites the system of magnetic moments
shown by red arrows. This additional dissipation results in 
current peaks in the I-V characteristics in magnetic superconductors.
 \label{Fig1} 
 } 
 \end{figure}

We use the quasistatic approach assuming that the space structure of the
magnetic field is the same as in the static vortex lattice, but the
field moves in the same way as vortex lattice does. Thus all
quantities describing the moving vortex lattice, i.e the magnetic field
and supercurrents, have
the dependence on the coordinates and time in the combination 
$({\bf r}-{\bf v}t)$. In the field interval $B\ll H_{c2}$ the magnetic
field should be found from the London equations \cite{Brandt,Kogan,bul} 
\begin{eqnarray}
&&{\rm curl}~{\bf B}=\frac{4\pi}{c}{\bf j}_s+4\pi{\rm curl}~{\bf M},  \label{B}\\
&&{\bf j}_s=\frac{c\Phi_0}{8\pi^2\lambda_{\perp}^2}
\left(\nabla\phi-\frac{2\pi}{\Phi_0}{\bf A}\right), \ \  {\bf B}=
{\rm curl}~{\bf A}, \label{A}\\
&&{\rm curl}~\nabla\phi=\sum_n2\pi\delta({\bf r}-{\bf r}_{n}),
\label{phi}
\end{eqnarray}
where ${\bf j}_s$ is the supercurrent, ${\bf A}$ is the vector
potential, $\phi$ is the phase of the superconducting order parameter,
 ${\bf M}$ is the local magnetization, 
$\Phi_0$ is the flux quantum and $\lambda_{\perp}=\lambda_x=\lambda_y$
is the London
penetration length for currents in the $(x,y)$ plane in the absence of
magnetic moments. Further, ${\bf r}_n(t)={\bf r}_n(0)+{\bf v}t$
are the coordinates of vortices and ${\bf r}_n(0)$
form a regular vortex lattice.
From Eqs.~(\ref{B})-(\ref{phi}) we obtain 
\begin{equation}
{\rm curl}~{\rm curl}~({\bf B}-4\pi{\bf M})+\frac{1}{\lambda_{\perp}^{2}}{\bf B}=
\frac{\Phi_0}{\lambda_{\perp}^{2}}\sum_n\delta({\bf r}-{\bf
r}_{n}).
\label{B1}
\end{equation}
To relate the Fourier components of $M_z({\bf r},t)\equiv M$ and $B_z({\bf
r},t)\equiv B$ 
we use the linear response approximation in which 
supercurrents induce the ``external'' magnetic field,
\begin{equation}
H({\bf k},\omega)=B({\bf k},\omega)-4\pi M({\bf k},\omega), 
\label{HBM}
\end{equation}
acting on the magnetic moments, where 
$M({\bf k},\omega)=\chi({\bf k},\omega)H({\bf k},\omega)$ and 
$\chi({\bf k},\omega)\equiv\chi_{zz}({\bf k},\omega)$ is the
susceptibility of the magnetic system.  
This approach is valid for the magnetization harmonics $g_x\neq 0$ satisfying
the condition 
\begin{equation}
|M({\bf g},g_xv)|^2/(\mu n_M)^2\ll 1. 
\label{Mg}
\end{equation}
For
an antiferromagnet with two sublattices the magnetic susceptibility is
given \cite{bul1,Levy,White} by
\begin{equation}
\chi({\bf k},\omega)=\frac{\omega_M\omega_s({\bf k})}
{\omega_s^2({\bf k})-\omega^2-i\omega\nu_s} \ .
\end{equation}
Here 
$\omega_M=\mu^2n_M/(2\hbar)$ at $\mu B\ll k_BT_M$, the density of
magnetic ions is $n_M$ and their magnetic moment is $\mu$,
$\omega_s({\bf k})$ is the magnetically active spin wave dispersion 
renormalized by the
superconductivity  \cite{bul1}, while
$\nu_s$ is the relaxation rate of spin
waves due to the interaction with phonons. 
Using Eqs.~(\ref{B1}) and (\ref{HBM}), we obtain for the
Fourier components $({\bf k}= {\bf g}\equiv2\pi(B_0/\Phi_0)^{1/2}(n,m,0); \omega = g_x v)$
of the magnetic field 
\begin{equation}\label{Bg}
 \left[1+\frac{\lambda_{\perp}^2{\bf k}^2}{1+4\pi\chi({\bf
k},\omega)}\right]
B({\bf k},\omega)=\sum_{\bf g}B_0\delta({\bf k}-{\bf g})\delta(\omega-g_xv)
\end{equation}
where $B_0$ is is the average induction
and $n,m$ are integer. From Eq.~(\ref{Bg}) we see that magnetic
moments renormalize the London penetration length so that the effective
penetration length in magnetic superconductors is given by \cite{bul}
\begin{equation}
\Lambda_{\perp}({\bf k},\omega)=\frac{\lambda_{\perp}}{[1+4\pi\chi({\bf k},
\omega)]^{1/2}}.
\end{equation}
Solving Eq.~(\ref{Bg}) we obtain the
Fourier components of the magnetic field $B$  and the ``external'' field
$H$  as
\begin{eqnarray}
&&B({\bf k},\omega)=B_0\frac{\delta({\bf k}-{\bf
g})\delta(\omega-g_xv)}
{1+\Lambda_{\perp}^2({\bf g},\omega)g^2}
, \\ 
&&H({\bf k},\omega)=B_0\frac{\delta({\bf k}-{\bf g})\delta(\omega-g_xv)}{1+4\pi\chi({\bf g},\omega)+
\lambda_{\perp}^2g^2}.
\end{eqnarray}
Thus the moving vortex lattice induces a spatially
periodic ac ``external'' magnetic field $h({\bf r},t)= H({\bf
r},t)-B_0$ along the $z$-axis
characterized by momenta ${\bf g}$ and washboard frequencies
$\omega=vg_x$.
At $\chi=0$ for $\lambda_{\perp}=1300$ \AA, typical for
borocarbides, the amplitude of the main harmonic, $n=1, m=0$, is about
20 G. The moving vortex lattice induces also an electric field ${\bf
E}=[{\bf v}\times{\bf B}]/c$ along the current direction. 

When the alternating magnetic field $h({\bf r},t)$ is not parallel to
the sublattice magnetization, it 
excites spin waves with   momenta ${\bf g}$ and   frequencies 
$\omega_s({\bf g})={\bf g}\cdot{\bf v}$
(this condition of resonance holds for any vortex lattice). 
Assuming that sublattice magnetizations are almost perpendicular to
the applied magnetic field, we obtain for the power per unit
volume transmitted from the vortex lattice to the magnetic system
the expression \cite{White} 
\begin{eqnarray}
&&{\cal P}_M=-\left\langle {\bf M}({\bf r},t)\cdot\frac{\partial{\bf h}({\bf r},t)}
{\partial t}\right\rangle= \nonumber \\
&&\sum_{\bf g}2g_xv|h({\bf g},g_xv)|^2{\rm Im}[\chi({\bf g},g_xv)],
\end{eqnarray}
where angular brackets denote time and space average. 

Now we are in a position to find the velocity of the vortex lattice at a
given transport current density $j$. For that we equate the power per unit
volume performed by the battery, $jE$, to the sum of the power
dissipated by quasiparticles, $\eta v^2$, and that transmitted to the
magnetic system, ${\cal P}_M$. Here $\eta$ is the viscous drag coefficient
due to quasiparticles in normal vortex cores. It is given by the
Bardeen-Stephen expression $\eta=B_0H_{c2}^*\sigma_n/c^2$, where $\sigma_n$
is the normal state conductivity, 
$H_{c2}^*=\Phi_0/(2\pi\xi_{\perp}^2)$ 
is the orbital upper critical field and $\xi_{\perp}$ is the
superconducting correlation length in the direction perpendicular to
the applied magnetic field. Taking into account that $E=vB_0/c$
and $\omega=vg_x=cEg_x/B_0$, we find $v$ and finally j-E (i.e., I-V)
characteristics in the intervals of $E$, where inequality
Eq.~(\ref{Mg}) is fulfilled:
\begin{equation}
j(E)=\frac{c^2\eta}{B_0^2}E+
\sum_{{\bf g}\neq 0}\frac{2g_xcB_0{\rm Im}[\chi({\bf
g},cEg_x/B_0)]}{|1+4\pi\chi({\bf
g},cEg_x/B_0)+{\bf g}^2\lambda_{\perp}^2|^2}.
 \label{jE}
\end{equation}
From this equation we see that the current density as a function of
$E$ has peaks corresponding to   resonances between the ac magnetic
field and spin waves, i.e when $\omega(n,m)=2\pi
v(B_0/\Phi_0)^{1/2}n$. 

Let us discuss the behavior of $j(E)$ near   resonances. We introduce
the frequency deviation $\Delta\omega=\omega_s({\bf g})-\omega$ such that
$\nu_s\ll\Delta\omega\ll\omega_s(n,m)$. Then we obtain $\chi({\bf g},\omega)
\approx \omega_M/(2\Delta\omega)$ and ${\rm Im}[\chi({\bf g},\omega)]\approx 
\omega_M\nu_s/(2\Delta\omega)^2$. We consider the interval of
frequency deviations 
$\Delta\omega$ where $\lambda_{\perp}^2g^2\gg 4\pi\chi({\bf
g},\omega)$. In this interval we estimate
\begin{equation}
\frac{M({\bf g},\omega)}{\mu n_M}\approx \frac{\mu
\Phi_0}{16\pi^2(n^2+m^2)\lambda_{\perp}^2\hbar\Delta\omega}.
\end{equation}
Due to the condition Eq.~(\ref{Mg}) our approach is valid for 
$\hbar\Delta\omega>\mu\Phi_0/(4\pi\lambda_{\perp})^2$. 
The ratio of the additional current caused by spin waves over
the current background is given as 
\begin{equation}
\frac{\Delta j(n,m)}{j}\approx
\frac{\omega_M\nu_s\Phi_0B_0}{8\pi^2\omega\eta(\Delta\omega)^2
\lambda_{\perp}^4} \frac{n^2}{(n^2+m^2)^2}.
\label{ratio}
\end{equation}
In the frequency interval $\hbar\Delta\omega>\mu\Phi_0/
(4\pi\lambda_{\perp})^2$ we obtain the inequality
\begin{equation}
\frac{\Delta j(n,m)}{j}<
\frac{16\pi^2\hbar  n_M\nu_sB_0}{\omega\eta\Phi_0}
 \frac{n^2}{(n^2+m^2)^2}.
\label{est}
\end{equation}
In magnetic insulators $\nu_s$ is typically of order $10^6$
s$^{-1}$. One can anticipate the same value in magnetic
superconducting crystals, as conducting electrons are gapped. For 
HoNi$_2$B$_2$C, taking $H_{c2}^*\approx 10$ T, $n_M=10^{22}$ cm$^{-3}$,
$\sigma_n=10^5$
(ohm$\cdot$ cm)$^{-1}$ at $\omega=10^{10}$ s$^{-1}$ we derive 
$\Delta j(n,m)/j<0.8n^2/(n^2+m^2)^2$. Thus, the peak $n=1, m=0$ is
observable even in the frequency interval where our linear response
approach is valid. 
Here the magnetic system deviates only slightly from 
equilibrium as energy is transformed further to phonon bath.

Closer to the resonance the linear response approach breaks down. 
Here the dominant contribution to dissipation comes from 
generation of spin
waves by vortices which leads to a strong deviation of the magnetic
system from equilibrium. 
For quantitative description of the j-E characteristics close to resonances 
the full dynamic approach for vortices and magnetic system is
necessary. 

Based on Eq.~(\ref{jE}) we see that measurements of the I-V
characteristics at different magnetic fields and currents may provide
information on the spin wave dispersion $\omega_s({\bf g})$. The
washboard frequency $\omega$ and the reciprocal vortex lattice vectors
${\bf g}$ may be changed independently by varying $B_0$ and ${\bf j}$,
 but an important question is
what are limitations on the variations of the magnetic field and the
current density. Momentum $k\sim
2\pi(B_0/\Phi_0)^{1/2}$ is of order $10^6$ cm$^{-1}$  in
fields $B_0 \leq 1$ T and increases as one 
approaches $H_{c2}$, but then harmonic amplitudes $h({\bf g},\omega)$
drop. Limitations
on frequency are due to limitations on the current density, which
should be lower than the depairing current density, and also should
not lead to excessive heating. From Eq.~(\ref{jE}), to reach
frequency $\omega$ one needs current density 
$j(\omega)\geq\sigma_n\omega H_{c2}^*/cg_x$ and the
electric field $E(\omega)=\omega B_0/cg_x$.
For $B_0=1$ T we obtain $j(\omega)\approx 10^7n^{-1}(\hbar\omega/1{\rm
K})$ A/cm$^2$ if the lowest harmonics are used, while for higher
harmonics higher frequencies may be reached. The depairing current
density for borocarbides is of order $10^7$ A/cm$^2$ and thus spin
waves with energies $\hbar\omega\lesssim 1$ K may be probed without
strong suppression of superconducting properties by the transport
current. For the dissipation power per unit volume, 
${\cal P}_{{\rm dis}}=jE\geq \sigma_n\omega^2\Phi_0H_{c2}^*/4\pi^2c^2$,
we estimate ${\cal P}_{{\rm dis}}\sim 10^8n^{-2}(\hbar\omega/1{\rm K})^2$
W/cm$^3$. To diminish heating the pulse technique may be used, as in I-V
measurements by Kunchur \cite{Kunchur}.

As only the low energy part of the spin wave spectrum may be probed by
I-V measurements, the important question is what is the minimum
energy of spin waves. For a Neel antiferromagnetic state there is
always magnetic
gap in the spectrum due to the magnetic anisotropy slightly renormalized by 
superconducting
screening of RKKY and dipole-dipole interactions \cite{bul1}. Neither
experimental nor theoretical information on the strength of magnetic
anisotropy or the structure of excitations in these materials is
available so far. Thus we cannot predict yet whether resonance
conditions for the lowest harmonics will be fulfilled in
borocarbides. However, we can anticipate that higher harmonics will be
effective in the case of weak pinning. 

Next we discuss highly anisotropic layered crystals like
SmLa$_{1-x}$Sr$_x$CuO$_{4-\delta}$. 
This material is
especially interesting because its T$^*$ structure leads to a
two-dimensional character of the magnetic system. Here magnetic
Sm$_2$O$_2$ and nonmagnetic La$_{2-x}$Sr$_x$O$_{2-\delta}$ layers
alternate in the barriers between the superconducting CuO$_2$
layers. The Josephson nature of the interlayer coupling in this
crystal has been confirmed by observation of the double Josephson
plasma resonance stemming from two layers in a unit cell
\cite{Shibata}. The specific heat measurements
\cite{Ronning} show that magnetic ordering is absent down to a temperature of
0.3 K and a magnetic gap, if any, lies below 0.3 K. They
reveal also a broad peak near the temperature 1 K and the height of this
peak indicates the presence of competing interactions that might be
described by the two-dimensional $J_1$-$J_2$ Heisenberg model with
$J_2/J_1>0.4$ \cite{Ronning,Mis}. Such a model has very complex
dynamics and contains a variety of transitions down to zero
temperature, making it an ideal testing ground for the theory of
quantum phase transitions. The most interesting part of the phase
diagram is in the region $0.4\lesssim J_2/J_1\lesssim 0.55$, where a
gapped phase without magnetic ordering is likely to be taking place. 
However, its characterization has been one of the most intriguing
puzzles of the physics of strongly correlated systems
\cite{Capriotti}. 

If the magnetic field is applied perpendicular to the layers (along
the c-axis), it induces pancake vortices which do not form a regular
lattice in   magnetic fields above 20 G as they order along the $c$-axis only
due to weak Josephson and magnetic interactions \cite{Blatter}. This
makes excitation of spin waves ineffective by moving vortex lattice 
induced by a perpendicular magnetic field. 
When a magnetic field is applied parallel to the layers (in the
$ab$-plane, along the $y$-axis), the situation
is drastically different, because now Josephson vortices
\cite{Kulik,Clem,bdmbi,Koshelev,KA} are induced. In high fields they
form a lattice which is quite regular in the $x$-direction (parallel to
the layers). Josephson vortices do not have normal cores and so only
thermally induced quasiparticles (or those near the nodes in the case
of d-wave pairing) cause dissipation. A weak interlayer tunneling
transport current, which
leads to vortex motion in the $x$-direction, cannot destroy
superconductivity and produces much less heating than in the case of 
isotropic or weakly anisotropic superconductors. 

The distribution of the magnetic field $B({\bf r})$ inside intrinsic Josephson
junctions is described by coupled finite-difference differential
equations for the phase difference $\varphi_n$ and for the magnetic
field $B_n$ inside the junction $n$ between layers $n$ and $n+1$ 
\cite{bdmbi,KA}. Accounting for the magnetization $M_n$ of ions inside
intrinsic Josephson junction $n$ we obtain equations for the
dimensionless variables $\varphi_n$, 
$b_n=B_n2\pi\lambda_{ab}\lambda_c/\Phi_0$, 
$m_n=M_n2\pi\lambda_{ab}\lambda_c/\Phi_0$ and $h_n=b_n-4\pi m_n$ :
\begin{eqnarray}
&&\frac{\partial^2\varphi_n}{\partial\tau^2}+
\nu_c\frac{\partial\varphi_n}{\partial\tau}+\sin\varphi_n-
\frac{\partial h_n}{\partial u}=0, \label{J}\\
&&\nabla_n^2h_n-\frac{b_n}{\ell^2}+\frac{\partial\varphi_n}{\partial u}+\nu_{ab}
\frac{\partial}{\partial\tau}\left(\frac{\partial\varphi_n}{\partial
u}-
\frac{b_n}{\ell^2}\right)=0, \nonumber
\end{eqnarray}
where $u=x/\lambda_J$, $\tau=t\omega_p$, $\lambda_J=\gamma s$,
$s$ is the interlayer distance, $\gamma=\lambda_c/\lambda_{ab}$ is the
anisotropy ratio, $\lambda_c$
and $\lambda_{ab}$ are the London penetration lengths for currents
along the $c$-axis and in the $ab$-plane, respectively, 
$\omega_p=c/(\lambda_c\sqrt{\epsilon_c})$ is the Josephson frequency, 
$\epsilon_c$ is the dielectric function along the $c$-axis,
$\nu_c=4\pi\sigma_c/(\omega_p\epsilon_c)$,
$\nu_{ab}=4\pi\sigma_{ab}/(\gamma^2\epsilon_c\omega_p)$, $\sigma_c$
and $\sigma_{ab}$ are quasiparticle conductivities along the $c$-axis
and in the $ab$-plane, respectively. Using linear response
approximation, $m_n=\chi b_n/(1+4\pi\chi)$, where $\chi\equiv \chi_{yy}$,
we see that $h_n=b_n/(1+4\pi\chi)$ satisfies the same equations as
$b_n$ at $\chi=0$,
but with the renormalized parameter $\tilde{\ell}^{-2}=
(1+4\pi\chi)\ell^{-2}$. For SmLa$_{1-x}$Sr$_x$CuO$_{4-\delta}$
we estimate $\tilde{\ell}^{-2}\ll 1$ because $\omega_M\approx
1.8\cdot 10^8$ s$^{-1}$, $\ell^2\approx 2\cdot 10^4$ at
$\mu=0.8\mu_B$, 
$n_M=5\cdot 10^{21}$ cm$^{-3}$ and $\lambda_{ab}\approx 2000$ \AA.  

In the following we consider large enough fields 
$B>B_J\equiv \Phi_0/(2\pi s\lambda_J)$. Then the Josephson
vortices fill all intrinsic junctions, overlap strongly and form a
regular triangular
lattice \cite{Clem,bdmbi,Koshelev,KA}. (An illustration is given in 
Ref.~\onlinecite{Clem}.) For SmLa$_{1-x}$Sr$_x$CuO$_{4-\delta}$
we have $\gamma\approx 500$, $\omega_p\approx 10^{12}$ s$^{-1}$ and
$B_J\approx 0.5$ T. In a Josephson system the washboard frequency is the
Josephson frequency $\omega=\omega_J=2eV/\hbar$, where $V$ is the voltage
between neighboring layers. For a triangular lattice at frequencies
and the magnetic fields satisfying the conditions
$\ell^2\gg(1+4\pi\chi)$ and 
$|2\tilde{\omega}-\tilde{b}|\gtrsim 1$, where $\tilde\omega=\omega/\omega_p$
and $\tilde{b}=B_0/B_J$, the solution of Eqs.~(\ref{J}) has the form 
\begin{eqnarray}
&&\varphi_n(u,\tau)\approx\tilde{\omega}\tau-\tilde{b}u+\pi
n+\frac{4\sin(\tilde{\omega}\tau-\tilde{b}u+\pi n)}
{4\tilde{\omega}^2-\tilde{b}^2}, \nonumber\\
&&h_n(u,\tau)\approx-h_0 \cos(\tilde{\omega}\tau-\tilde{b}u+\pi n), \ \ 
h_0 \approx \frac{\tilde{b}}{4\tilde{\omega}^2-\tilde{b}^2},
\nonumber\end{eqnarray}
where we neglected $\nu_c$ and $\nu_{ab}$. We estimate $h\equiv h_0 \Phi_0/(2\pi\lambda_{ab}^2\gamma)\approx
0.16$ G at $\omega=0.1\omega_p$ and $B=B_J$. Near the Eck resonance, 
$2\tilde{\omega}\approx\tilde{b}$, the amplitude of the magnetic field
$h$ is larger. For the reciprocal lattice vector we have ${\bf g}=(2\pi
sB/\Phi_0, 0, \pi/s)$. So 
$g_x=1/\lambda_J\approx 10^4$ cm$^{-1}$ at $B=B_J$. 

Assuming that sublattice magnetization is almost perpendicular to the
applied magnetic field 
 or that magnetic ordering is absent 
we obtain for the I-V characteristics
\begin{equation}\label{jLayered}
j(V)=\sigma_{{\rm eff}}\frac{V}{s}+\frac{esh^2}{\hbar}{\rm
Im}\left[\chi_{yy}\left({\bf g},\frac{2eV}{\hbar}\right)\right],
\end{equation}
where $\sigma_{{\rm eff}}=\sigma_c+2\sigma_{ab}B_J^2/(\gamma B)^2$
describes dissipation due to quasiparticles. 
At resonance  $\omega_J=\omega_s({\bf g})$, we estimate
$\Delta j/j\approx 2\pi^2c^2s^2h^2\omega_M/(\omega_J\sigma_{{\rm
eff}}\nu_s\Phi_0^2)$.
Estimating $\sigma _c \simeq 10^{-3} (\Omega $cm$)^{-1}$
and $\sigma _{ab} \simeq 4 \cdot 10^{4} (\Omega $cm$)^{-1}$ as in BSCCO
and taking $s\simeq 12$ \AA \  as in Sm$_{2-x}$Ce$_{x}$CuO$_{4-\delta}$,
we obtain
 $\Delta j/j\approx 4$ and $|M({\bf g},\omega_J)|/(\mu
n_M)\approx 0.3$ at $\omega=10^{12}$ s$^{-1}$ and $B=B_J$ and bigger
values near the Eck resonance. Certainly, such frequencies are
sufficient to probe almost complete spectrum in 
SmLa$_{1-x}$Sr$_x$CuO$_{4-\delta}$. 

In conclusion, we propose to probe low-frequency magnetic excitations
in magnetic superconductors by measuring I-V characteristics in the
mixed state with a moving vortex lattice. Coupling of such a lattice
to magnetic moments is due to an ac magnetic field which is inherent
to vortex motion. The energy interval of spin waves which can be
probed in isotropic and moderately anisotropic superconductors is
limited by the depairing current and heating. If
spin wave energies fall in this interval, they affect the vortex
motion strongly and should be easily seen in the I-V characteristics
as current peaks at corresponding voltages. Such an effect may be
observed in borocarbides if they
have spin waves with energies below 1 K. For highly anisotropic
layered superconductors in parallel magnetic fields, higher spin wave energies
may be probed by use of moving Josephson vortices. This is sufficient
to study almost complete spin wave spectrum in
SmLa$_{1-x}$Sr$_x$CuO$_{4-\delta}$ with exotic magnetic ordering.

The authors thank J. Clem, A. Buzdin, V. Kogan, A. Koshelev, C. Batista, D. Smith, R. Movshovich,
F. Ronning and S. Goupalov for helpful discussions. 
This work was supported by NNSA.

\end{document}